\documentclass[5p,times,authoryear]{elsarticle}

\biboptions{square,comma}
\setcitestyle{numbers}

\usepackage[colorlinks=true,linkcolor=blue,citecolor=blue,urlcolor=blue]{hyperref}
\usepackage{graphicx}
\usepackage{dcolumn}
\usepackage{bm}
\usepackage{color}
\usepackage{hyperref}
\usepackage{soul}
\usepackage{amsmath}
\usepackage{amssymb}

\begin{document}

\title{Emulators for scarce and noisy data:\\application to auxiliary field diffusion Monte Carlo for the deuteron}

\author[1,2]{Rahul~Somasundaram}
\ead{rsomasundaram@lanl.gov}
\author[3]{Cassandra~L.~Armstrong}
\author[4,5]{Pablo~Giuliani}
\author[4]{Kyle Godbey}
\author[2]{Stefano Gandolfi}
\author[2]{Ingo Tews}

\address[1]{Department of Physics, Syracuse University, Syracuse, NY 13244, USA}
\address[2]{Theoretical Division, Los Alamos National Laboratory, Los Alamos, NM 87545, USA}
\address[3]{Intelligence and Space Research Division, Los Alamos National Laboratory, Los Alamos, NM 87545, USA}
\address[4]{Facility for Rare Isotope Beams, Michigan State University, East Lansing, Michigan 48824, USA}
\address[5]{Department of Statistics and Probability, Michigan State University, East Lansing, Michigan 48824, USA}

\date{\today}

\begin{abstract}
The validation, verification, and uncertainty quantification of computationally expensive theoretical models of quantum many-body systems require the construction of fast and accurate emulators. 
In this work, we develop emulators for auxiliary field diffusion Monte Carlo (AFDMC), a powerful many-body method for nuclear systems.
We introduce a reduced-basis method (RBM) emulator for AFDMC and study it in the simple case of the deuteron.
Furthermore, we compare our RBM emulator with the recently proposed parametric matrix model (PMM) that combines elements of RBMs with machine learning. 
We contrast these two approaches with a traditional Gaussian Process emulator. 
All three emulators constructed here are based on a very limited set of 5 training points, as expected for realistic AFDMC calculations, but validated against $\mathcal{O}(10^3)$ exact solutions.
We find that the PMM, with emulator errors of only $\approx 0.1 \%$ and speed-up factors of $\approx 10^7$, outperforms our implementation of the other two emulators when applied to AFDMC. 
\end{abstract}

\maketitle

\section{Introduction}

In the next years, an explosion of new data from laboratory experiments, such as the Facility for Rare Isotope Beams (FRIB), and multi-messenger observations of neutron stars and their mergers~\cite{Koehn:2024set,NANOGrav:2019jur,LIGOScientific:2017ync,LIGOScientific:2017vwq,Riley:2019yda,Riley:2021pdl,Miller:2019cac,Miller:2021qha} will provide exciting new information for nuclear physics.
To robustly analyze the information provided during this data-rich era, reliable theoretical approaches with well-quantified uncertainties are key.
These approaches can then be employed in statistical tools based on Bayesian inference~\cite{Koehn:2024set,Pang:2022rzc,Raaijmakers:2019qny,Farr:2021fyc,phillips2021get}. 
Quantum Monte Carlo (QMC) methods~\cite{Carlson:2014vla}, such as auxiliary field diffusion Monte Carlo (AFDMC)~\cite{Schmidt:1999lik}, combined with interactions from chiral effective field theory (EFT)~\cite{Gezerlis:2013ipa,Gezerlis:2014zia,Tews:2015ufa,Lynn:2015jua,Somasundaram:2023sup} are some of the most universal and reliable nuclear many-body approaches used in the community.
QMC algorithms can be applied to both atomic nuclei and nuclear matter using the same input interactions, which enables us to straightforwardly connect nuclear experiments with astrophysical observations. 
QMC methods are also very accurate and precise and provide non-perturbative, virtually exact solutions to the Schr{\"o}dinger equation~\cite{Carlson:2014vla,Lonardoni:2017hgs}. 
However, these benefits incur a large computational cost, of the order of several $100,000$~CPU-h per typical simulation.

The Bayesian approaches necessary for the upcoming data-rich era typically require a large number of model evaluations across a broad parameter space, rendering their application to expensive numerical approaches, such as QMC, prohibitively expensive. 
Emulators, i.e., algorithms that mimic the behavior of a high-fidelity model at a fraction of its computational cost, have been proposed to circumvent this problem~\cite{Bonilla:2022rph,Melendez:2022kid,Duguet:2023wuh,Frame:2017fah,Konig:2019adq,Lay:2023boz,surer2022uncertainty,rose2024,drischler2023buqeye}. 
They can broadly be classified into two categories: intrusive and non-intrusive~\cite{Duguet:2023wuh}. 
Non-intrusive emulators are usually trained only on the inputs and outputs of the high-fidelity model and are agnostic to the underlying physics. 
Common examples include Gaussian process (GP) regression and artificial neural networks~\cite{GP_book,surer2022uncertainty,Lay:2023boz}.
Intrusive emulators, on the other hand, usually work with high-dimensional structures (such as wave functions), and respect certain physical aspects of the underlying equations or dynamics of the system.
Examples are reduced order models, such as reduced basis methods (RBMs)~\cite{hesthaven2016certified,Bonilla:2022rph,PG_book,J_1984,Frame:2017fah,Konig:2019adq,rose2024,drischler2023buqeye}, or dynamic mode decomposition and sparse identification of nonlinear dynamics (SINDy)~\cite{brunton2019data}.
While significant effort has been devoted to developing RBM-based emulators for some many-body techniques~\cite{Konig:2019adq,Ekstrom:2019lss,Jiang:2022tzf,Jiang:2022oba,Duguet:2023wuh}, emulators for QMC methods are in their earlier stages~\cite{Frame:2019jsw,Sarkar:2023qjn}. 
The primary obstacle to developing such emulators is the inability to calculate inner products between eigenstates of different Hamiltonians - something that is crucial in the framework of RBMs for quantum systems. 
Recently Ref.~\cite{Sarkar:2023qjn} proposed the floating block method as a possible solution to this obstacle, and it was successfully applied to lattice Monte Carlo calculations of light nuclei.

\begin{figure}[t]
    \centering
    \includegraphics[width=0.94\columnwidth]{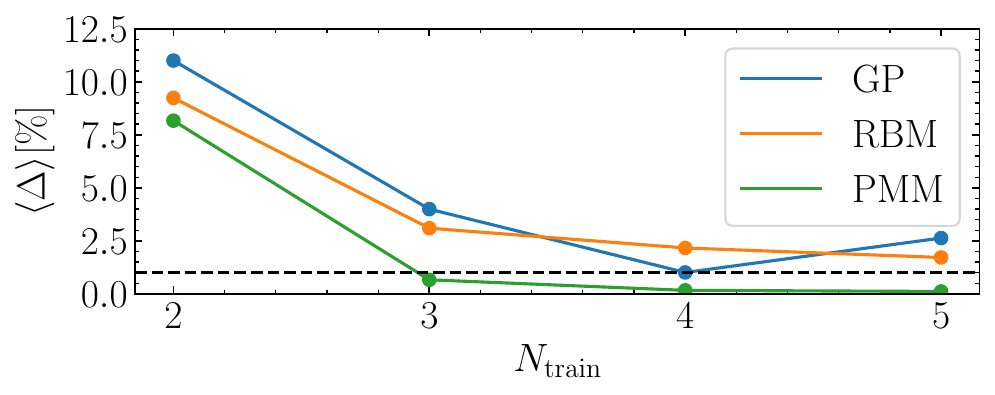}\\
    \vspace{-1.5mm}
    \includegraphics[width=0.94\columnwidth]{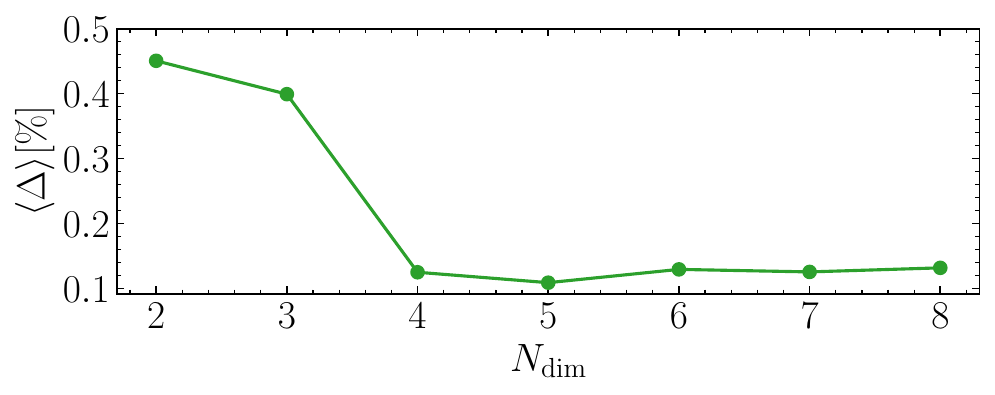}
    \caption{
    Top:  Averaged percentage error $\langle \Delta \rangle$ for the three emulators with respect to $N_{\rm train}$. 
    We indicate a 1\% error which is comparable to typical QMC errors~\cite{Carlson:2014vla} with the dashed black line.
    For the PMM, the matrix dimension $N_\text{dim}=5$.
    Bottom: $\langle \Delta \rangle$ of the PMM with respect to $N_\text{dim}$ using 5 AFDMC deuteron calculations as training points. 
    $\langle \Delta \rangle$ is computed by averaging over 1000 validation errors, each corresponding to a different validation sample.}
    \label{fig:n_train}
\end{figure}

\begin{figure}
    \centering
    \includegraphics[width=0.99\columnwidth]{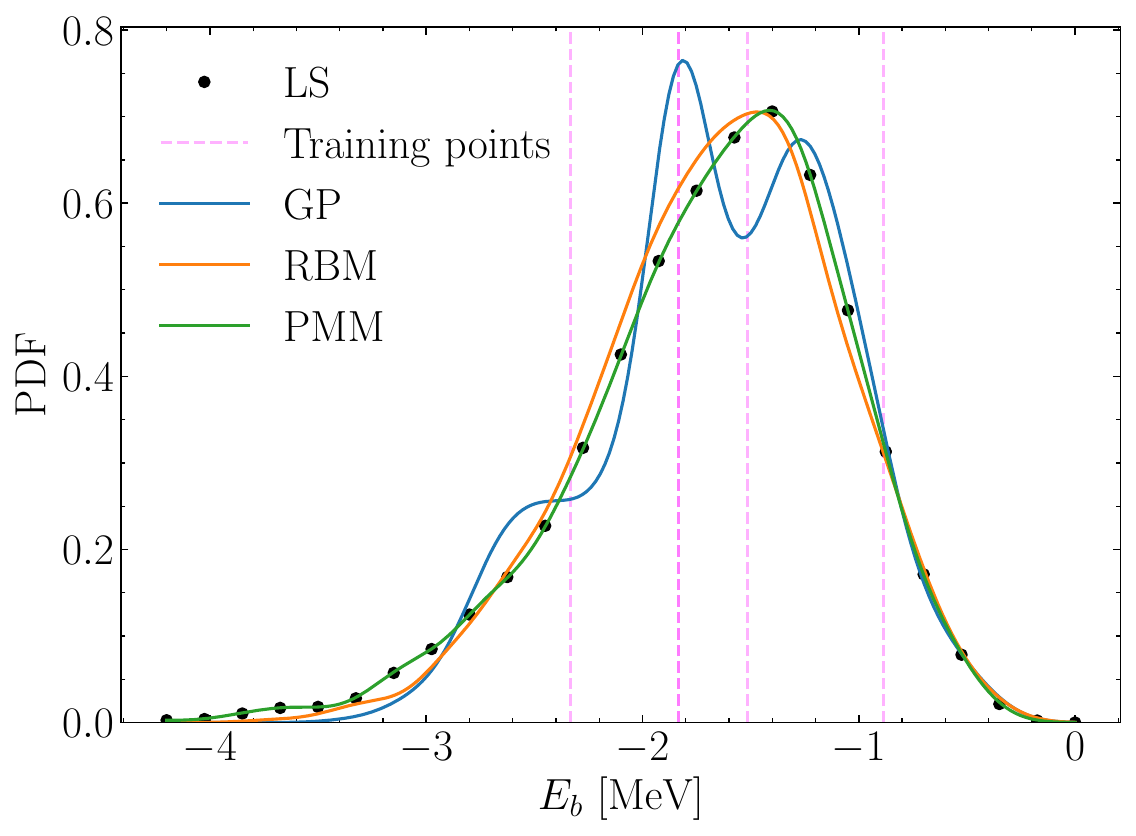}
    \caption{PDFs for the deuteron binding energy $E_b$ calculated with the three emulators and the exact LS solver.
    The PDFs result from a Gaussian kernel density estimator calibrated to the predictions obtained from the emulators and the exact LS solver. }
    \label{fig:PDF}
\end{figure}

In this letter, we develop novel emulators for AFDMC calculations of the deuteron.
First, we develop an intrusive RBM-based emulator using the Petrov-Galerkin projection method~\cite{Bonilla:2022rph,Melendez:2022kid} which circumvents the need to compute overlaps between exact AFDMC eigenstates of different Hamiltonians while maintaining the fully intrusive nature of the emulator. 
Second, we implement the parametric matrix model (PMM)~\cite{Cook:2024toj}, a machine learning algorithm that combines elements of both intrusive and non-intrusive emulators. 
Last, we compare these emulation methods to a traditional, non-intrusive GP emulator.
While AFDMC calculations of the deuteron are relatively less expensive than those of heavier systems, we expect to be limited by a  small set of training data, $N_{\rm train}\approx 5-10$, when developing AFDMC emulators in general.
Therefore, in this letter we require that our emulators achieve errors of a few percent when validated against $\mathcal{O}(10^3)$ exact solutions despite using a limited number of training data.
We find that our intrusive RBM and PMM emulators outperform the non-intrusive GP emulator, see Figs.~\ref{fig:n_train} and~\ref{fig:PDF}. 
Furthermore, the PMM performs best, with an average emulation error of only $\approx 0.1 \%$ but with a gain in speed of up to $\approx 10^7$ with respect to the AFDMC method. 
Our emulators enable us to propagate uncertainties directly from the Hamiltonian to observables, without the need of {\it a posteriori} approaches.

\section{Methods}

We employ the local chiral EFT two-nucleon interactions of Refs.~\cite{Gezerlis:2013ipa,Gezerlis:2014zia,Somasundaram:2023sup}. 
These interactions were calibrated to neutron-proton phase shifts in Ref.~\cite{Somasundaram:2023sup} using Bayesian inference, which results in posterior distributions over the low energy couplings (LECs). 
The nuclear Hamiltonian is then given by $H(\Vec{c})$, where $\Vec{c}$ is a set of control parameters, i.e., the LECs. 
We employ these interactions at cutoff $R_0=0.9$~fm in AFDMC, a continuum diffusion Monte Carlo (DMC) code~\cite{Schmidt:1999lik,Carlson:2014vla,Lynn:2019rdt}.
Starting from a trial wave function for a specific system, AFDMC performs an evolution in imaginary time to project out the ground state of the system.
All integrals appearing in this evolution are solved by means of Monte Carlo techniques. 
In addition, AFDMC achieves a better polynomial scaling with nucleon number than other DMC algorithms by linearizing spin-isospin states using a Hubbard-Stratonovich transformation~\cite{Schmidt:1999lik}.
However, as with all QMC algorithms, AFDMC results carry statistical noise.
In this work, we focus on the deuteron -the simplest atomic nucleus- since we can obtain exact solutions for validating our various emulation strategies in a reasonable time by solving the homogeneous part of the Lippmann–Schwinger (LS) equation~\cite{Epelbaum:2004fk}. 
All training and validation samples are drawn from the posterior distributions of $\Vec{c}$.  
We have checked that the deuteron energies calculated by the AFDMC and LS solvers agree to within $0.1 \%$.

To construct an RBM, one typically obtains $N_\text{train}$ high-fidelity solutions, i.e. ground states $\{| \psi \rangle \}_{j=1}^{N_\text{train}}$, corresponding to $\{\Vec{c}\}_{j=1}^{N_\text{train}}$, and then constructs a reduced basis of size $n\leq N_\text{train}$ informed by these solutions~\cite{Bonilla:2022rph}.
In this work we choose the reduced basis directly as the training solutions, a choice known as a Lagrange basis~\cite{quarteroni2011certified}. 
Then, for $\Vec{c}$ not in the training set, one determines the ground state $|\psi \rangle$ within the subspace spanned by $\{| \psi \rangle \}_{j=1}^{N_\text{train}}$, i.e., we impose $|\psi \rangle \approx \sum_j a_j |\psi_j \rangle$. 
Under this ansatz, the Schr{\"o}dinger equation becomes
\begin{equation}
    \sum_j  H  |\psi_j \rangle a_j = E \sum_j   |\psi_j \rangle a_j\,.
    \label{eq:general}
\end{equation}
This equation is then projected onto a subspace spanned by ``test" or ``projecting" functions~\cite{Bonilla:2022rph}.
In several cases, these are chosen as $\langle \psi_i |$, thereby casting Eq.~\eqref{eq:general} as a generalized eigenvalue problem for the matrix $M_{ij} \equiv \langle \psi_i |  H  |\psi_j \rangle$ with the norm matrix $N_{ij} = \langle \psi_i   |\psi_j \rangle$. 
In QMC approaches, these overlaps are dominated by stochastic noise~\cite{Carlson:2014vla}, see however Ref.~\cite{Sarkar:2023qjn} for a solution. 

Here, we circumvent this problem by using a novel application of the Petrov-Galerkin projection method~\cite{Bonilla:2022rph,PG_book,J_1984}.
We choose the projecting functions used to act on Eq.~\eqref{eq:general} from a different subspace as that spanned by $\{| \psi \rangle \}_{j=1}^{N_\text{train}}$. 
Considering a different $N_\text{train}$ dimensional subspace spanned by $\{| \phi \rangle \}_{i=1}^{N_\text{train}}$, the non-orthogonal projection results in a generalized eigenvalue problem for the matrix $\Tilde{M}_{ij} \equiv \langle \phi_i |  H  |\psi_j \rangle$ with the corresponding norm matrix $\Tilde{N}_{ij} \equiv \langle \phi_i  |\psi_j \rangle$. 
Here, we choose the $N_\text{train}$ trial wave functions that are used as initial conditions for the imaginary time evolution performed in AFDMC~\cite{Carlson:2014vla} as projecting functions.
Note that although an exact solution for the deuteron is possible, the trial wavefunctions employed here are not chosen to be exact, on purpose, in order to mimic the case of heavier systems. In particular, our trial wavefunctions are built using single-particle orbitals and correlations that do not satisfy the correct asymptotic behavior. However, the trial wave functions are optimized using variational Monte Carlo~\cite{Carlson:2014vla} and therefore have large overlap with the fully evolved AFDMC states $\{| \psi \rangle \}_{j=1}^{N_\text{train}}$. 
All required overlaps and matrix elements can be readily computed in AFDMC using the trial wave functions on one side and the evolved configurations on the other side.
Upon solving this generalized eigenvalue problem for a non-Hermitian matrix, we discard the complex eigenvalues and then take the smallest real eigenvalue to be our physical ground state energy. 
Unlike Refs.~\cite{Jiang:2022tzf,Jiang:2022oba}, we have not encountered spurious bound states that may arise due to the non-Hermiticity of the generalized eigenvalue problem.   
In a few cases, the two smallest real eigenvalues are very close to each other, less than $0.1$~MeV apart. 
Then, we take the average of the two eigenvalues which does not significantly change our results. 

In addition to this RBM emulator, we implement the PMM of Ref.~\cite{Cook:2024toj} for AFDMC. 
Inspired by the reduced equations obtained from RBMs~\cite{drnuclear}, we assume that the ground state energy of $H(\Vec{c})$ can be well approximated by the lowest eigenvalue of a matrix given by
\begin{equation}
    A(\Vec{c}) = A_0 + \sum_i c_i A_i 
    \label{eq:PMM}\,.
\end{equation}
Here, the $c_i$ are the LECs of the chiral Hamiltonian and we have used the fact that the LECs are affine. 
In contrast to the traditional RBM discussed above, we do not compute the matrix elements of $A_i$ from AFDMC wave functions. 
Instead, we take a data-driven approach and infer the matrices by fitting the lowest eigenvalue of $A$ to AFDMC results for the deuteron binding energy $E_b$ for different $\Vec{c}$.
A global optimizer, such as the basin-hopping algorithm~\cite{wales1997global}, helps find a suitable set of matrices -not necessarily unique- that reproduce the desired dynamics. 
We impose that the $A_i$ are real, symmetric matrices and that $A_0$ is diagonal~\cite{Cook:2024toj}.
The dimensionality of the matrices $A_i$, $N_\text{dim}$, is a hyperparameter of the emulator. 

We compare our RBM and PMM emulators with a non-intrusive GP emulator. 
For the GP kernel, we use a linear combination of the Matérn and the dot-product kernels~\cite{GP_book}, since many of the other standard kernels (such as the radial basis function) performed poorly in comparison. 
The optimization of the kernel hyperparameters was performed using the python package $\textsc{scikit-learn}$. 

\begin{figure}
    \centering
    \includegraphics[width=0.99\columnwidth]{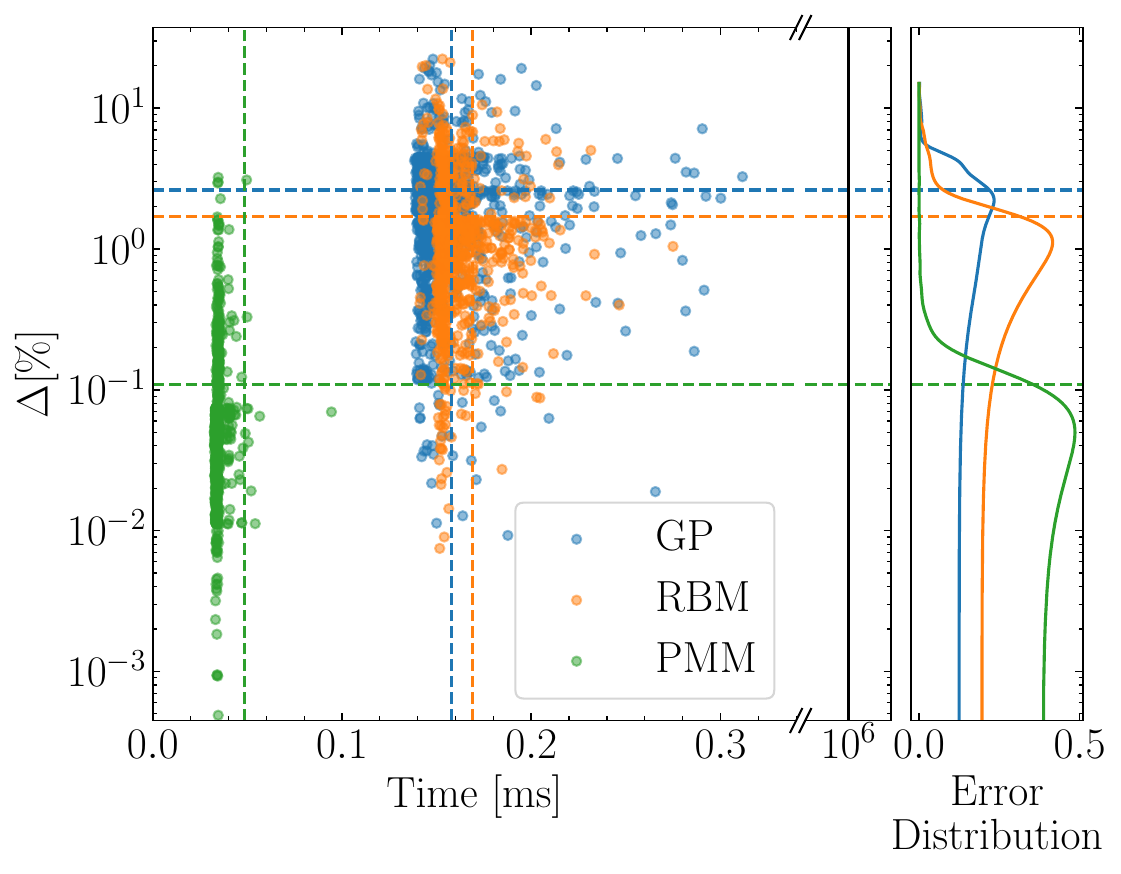}
    \caption{Computational accuracy vs. time plot~\cite{rose2024} for each model evaluation for our three emulators.
    Each dot corresponds to a validation sample.
    The dashed lines indicate the averages over all validation samples.
    The total CPU time required to perform a full AFDMC calculation is $\mathcal{O}(10^3)$s, which is indicated by the solid black vertical line. In the right panel, we show the distributions of the errors for our three emulators.
    }
    \label{fig:CAT_plot}
\end{figure}

\section{Results}

We first discuss our results for leading-order (LO) chiral EFT interactions.
At LO, we draw 5 samples from the posterior distribution on the single spectral LEC in the deuteron channel, $C_{^3\text{S}_1}$, obtained in Ref.~\cite{Somasundaram:2023sup}. 
These 5 interactions are then used in AFDMC calculations of $E_b$ as well as the various overlaps required for our RBM emulator. 
This set of 5 samples constitutes our training set. 
We draw a different set of 1000 samples from the same posterior distribution on $C_{^3\text{S}_1}$ and compute $E_b$ for each interaction by solving the LS equation. 
These calculations are used for validation.
For the training samples, the AFDMC and LS results agree on the sub-percent level.

In Fig~\ref{fig:n_train}, we show the performance of the three emulators as a function of $N_\text{train}$.
The averaged percentage error is given as:
\begin{equation}
  \langle  \Delta \rangle [\%]= \frac{1}{N} \sum_i^N \bigg| \frac{E_{b,i}^{\text{pred}} - E_{b,i}^{\text{LS}}}{E_{b,i}^{\text{LS}}} \bigg| \times 100\,,
\end{equation}
where $E_{b,i}^{\text{pred}}$ ($E_{b,i}^{\text{LS}}$) is the deuteron binding energy predicted by the emulator (LS solver) for sample $i$, and the sum is over the $N$ validation samples. 
The PMM clearly performs better than the other two methods, achieving sub-percentage emulation errors for $N_\text{train} \geq 3$. 
We have checked that adding terms non-linear in $c_i$ to Eq.~\eqref{eq:PMM} does not generally improve the accuracy of the PMM.
We have also studied the performance of the PMM with respect to $N_\text{dim}$ and found that it does not improve for $N_\text{dim}\gtrapprox 5$.
Therefore, we will use $N_\text{dim} = 5$ at leading order (LO) in the rest of this letter. 
We note that the validation uncertainty depends on the choice of the training set. 
To study this, we have generated a large ensemble of PMMs with $N_\text{dim}=5$, but trained with different sets of 5 training points. 
The maximum error over all these PMMs was found to be $\approx 0.3\%$, which is consistent with the PMM results of Fig.~\ref{fig:n_train}. 
This demonstrates the robustness of the PMM with respect to changes in the composition of the training set.
Our RBM emulator performs worse than the PMM.
With $N_\text{train} = 5$, our RBM achieves an average percentage error of $1.7 \%$, which is comparable to other RBMs reported in the literature (see Ref.~\cite{Ekstrom:2019lss}, the Lagrange Basis results in Table I of Ref.~\cite{Bonilla:2022rph}, and Table I of Ref.~\cite{Sarkar:2023qjn}). 
Finally, we find that the GP emulator generally performs worst.

Next, we fix $N_\text{train} = 5$ and calculate $E_b$ for the 1000 samples in our validation set. 
This results in emulated posterior probability density functions (PDF) on the binding energy shown in Fig.~\ref{fig:PDF}. 
The non-Gaussian PDF estimated from the exact LS solver is shown as reference. 
We see that the PDFs obtained from the PMM and the exact LS solver are virtually indistinguishable.
The Kullback–Leibler (KL) divergence~\cite{kullback1951information} between the two distributions is $5 \times 10^{-5}$. 
On the other hand, the KL divergence of the PDF predicted by our RBM (GP) emulator with respect to the LS solution is 0.008 (0.03). 
We find that our RBM emulator performs comparably well and captures the general shape of the PDF which results only in small differences for the percentiles of the PDF. 
In contrast, the PDF obtained from the GP emulator shows significant deviations from the exact one. 
This demonstrates the benefit of building intrusive or hybrid emulators when the training data set is limited.

Figure~\ref{fig:CAT_plot} depicts the percentage error for the emulators as a function of the computation time for each validation sample.
We find that the PMM, in addition to being more accurate than the other two methods, is faster by almost an order of magnitude. 
This is because solving the generalized eigenvalue problem for the RBM is slower than solving the regular eigenvalue problem for the PMM.
The spread in the errors provides an estimate of the outliers present in the data. 
We find that the biggest outlier for the PMM has an error of $\approx 3 \%$. 
On the other hand, both our RBM and the GP contain outliers with errors as large as $20 \%$. However, as can be seen from the distribution of errors shown in the right panel, our RBM contains significantly fewer outliers than the GP. 
Also, note that the emulators for QMC of Ref.~\cite{Sarkar:2023qjn} result in even larger outliers.
We found that the AFDMC calculations of the overlaps $\langle \phi_i  |\psi_j \rangle$ are highly correlated throughout the imaginary-time evolution.
As a consequence, our results carry larger stochastic noise. 
We expect that further investigating mitigation strategies for this AFDMC noise as well as improving the choice of the trial wave functions will lead to an improvement in the RBM performance. 
These lines of research will be important as we move to more complex systems since, among the emulators we built here, the fully intrusive RBM is the only one that has access to the underlying high dimensional wave functions. 

\begin{figure}[t]
    \centering
    \includegraphics[width=0.99\columnwidth]{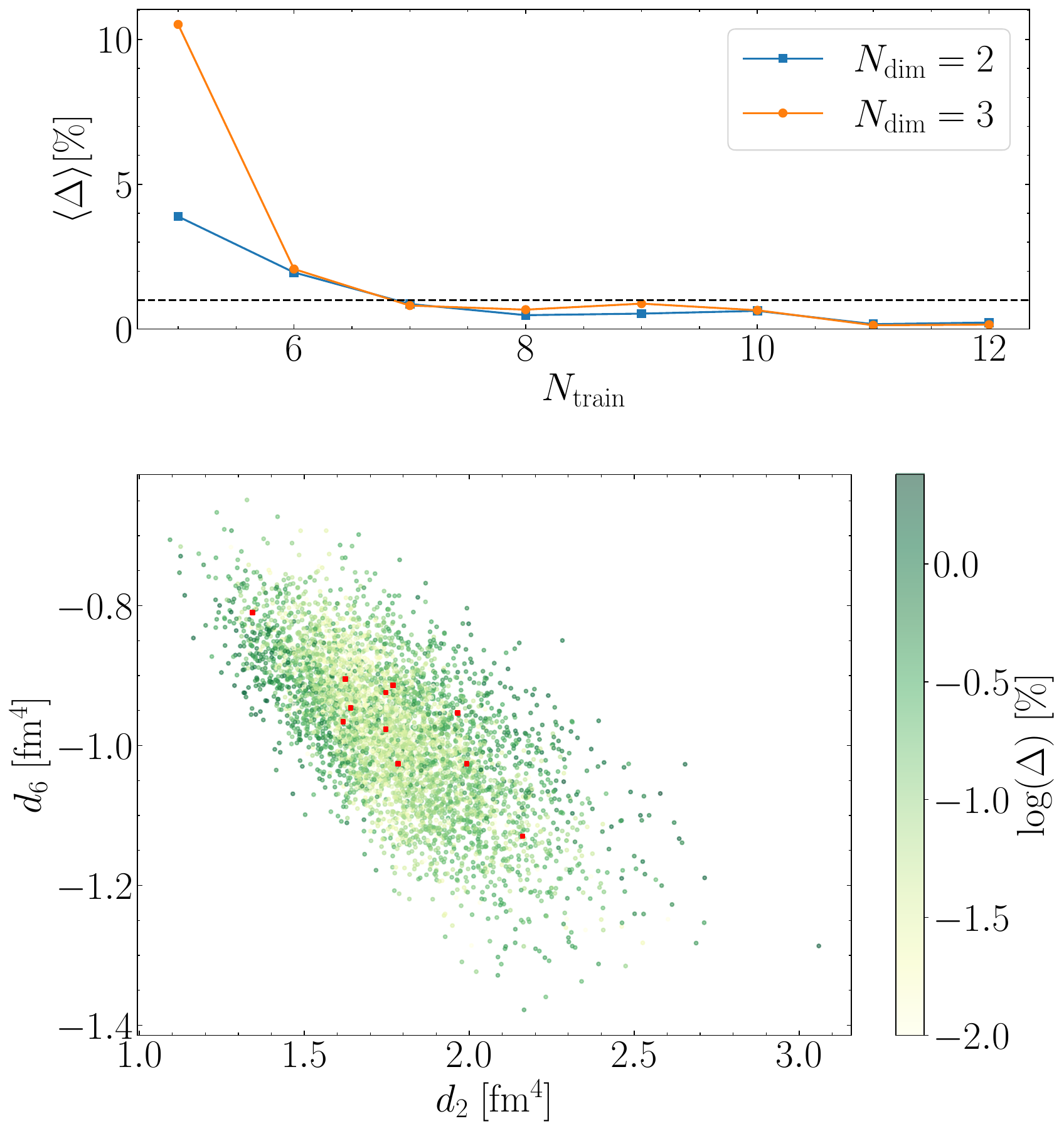}
    \caption{Top: Behaviour of the PMM for local N$^2$LO interactions with respect to the number of training samples. 
    The average percentage error is calculated by validating against 5000 samples.
    Bottom: Relative error $\Delta$ of the PMM emulator for the 5000 validation samples with respect to the LECs $d_2$ and $d_6$. 
    The values of $d_2$ and $d_6$ for the training points are shown are red.}
    \label{fig:d_2_vs_d_6}
\end{figure}

So far, we have considered only LO interactions that contain only one spectral LEC in the deuteron channel. 
We now study a larger parameter space by employing interactions at N$^2$LO.
At this order, four independent spectral LECs contribute to the deuteron. 
As we found the PMM to perform best at LO, we focus on the PMM for N$^2$LO and draw up to $12$ samples from the posteriors on the LECs calculated in Ref.~\cite{Somasundaram:2023sup} for training. 
Since the PMM requires only the energy of the deuteron for training, we generate the binding energies using the LS solver.
For validation, we now use 5000 samples drawn from the same posterior. 
We show our results for PMMs with $N_\text{dim} = 2$ and $N_\text{dim} = 3$ in Fig.~\ref{fig:d_2_vs_d_6} (top). 
We find that both PMMs achieve sub-percentage accuracies for $N_\text{train} \geq 7$, even in a 4 dimensional parameter space. 
For the PMM with $N_\text{dim} = 2$ and $N_\text{train} = 11$, the average error is $0.2 \%$ with the largest outlier having an error of $2.4 \%$. 
Similar to the results of Fig.~\ref{fig:PDF}, the PDF on the deuteron binding energy calculated with this PMM has a vanishingly small KL divergence of $3 \times 10^{-4}$ with respect to the PDF evaluated with the LS solver. 
Finally, we see that the PMM with $N_\text{dim} = 3$ performs slightly worse than the one with $N_\text{dim} = 2$ for $N_\text{train} < 10$. 
This is due to the fact that it is harder to infer larger matrices $A_i$~\eqref{eq:PMM} when the training set is limited and the number of parameters increases.

For the PMM with $N_{\rm dim}=2$ and $N_{\rm train}=11$, we further study the relative errors within the parameter space spanned by the 4 relevant spectral LECs. 
In Fig.~\ref{fig:d_2_vs_d_6} (bottom), we show the relative error $\Delta$ of the PMM for all of our validation samples.
We limit ourselves to a projection onto the plane spanned by the tensor coupling $d_2$ and the spin-orbit coupling $d_6$ but other LEC choices lead to similar results.
We find that the PMM both interpolates and extrapolates well even when the number of unknowns (14) is larger than the number of training points (11). 
The PMM is still very effective for parameter values far from those of the training points and we see some regions of particularly good performance outside of the range of training points.
Note that the same point in $d_2-d_6$ space can have different $\Delta$ due to different values of the other LECs.
We found that generally there can be an upper bound on the predictions of PMMs that depends on the LECs.
This bound is generally more important at low $N_{\rm dim}$. 
However, in our case, this bound is positive and does not impact our results.

\section{Conclusion}

We have developed three emulators for AFDMC calculations of the deuteron: a Petrov-Galerkin RBM-based emulator, a PMM, and a GP emulator. 
As shown in Fig.~\ref{fig:PDF}, we found the RBM and PMM to be generally superior to a traditional GP regression, both performing well despite being trained on a very limited data set.
We expect that the RBM performance can be further optimized in the future by both reducing the inherent noise in the reduced basis as well as for the selection of the projecting functions.
This is an important goal because these intrusive emulators give access to the complete wave functions of the many-body system, which allows us to easily compute other matrix elements of interest. 
On the other hand, we have demonstrated that PMMs are already a good choice for the purposes of emulating the ground-state energy of these systems. 
In addition to their performance, the PMM's almost straightforward set-up establishes them as a valuable new emulation tool for a broad class of systems. 
While the focus of this letter has been on the emulation of AFDMC calculations of the deuteron, we have tested the PMM emulator for Faddeev calculations of the triton binding energy using the same local N$^2$LO interactions as above, but supplemented with the leading chiral three-nucleon forces.
We have found that the PMM performs remarkably well, with a sub-percent validation uncertainty, demonstrating its potential to scale to more complex nuclear systems and interactions. 
Furthermore, we have applied the PMM emulator to AFDMC calculations of pure neutron matter recently~\cite{Armstrong:2025tza} and found the emulator to work very well, with an emulator uncertainty of order of 1\%. 
Finally, we note that the PMM approach is agnostic to the employed many-body approach and is not restricted to AFDMC. Our findings in this paper are therefore also relevant for the construction of emulators for other many-body methods as well.
Therefore, we believe that the calculations and tools developed in this work will enable novel applications of chiral interactions and QMC methods, such as their implementation in data analyses pipelines used to interpret multi-messenger neutron star observations. \\

\section*{Acknowledgements}
\begin{sloppypar}
We thank J.~Carlson, C.~Drischler, R.~Furnstahl, B.~Reed, and R.~Weiss for insightful discussions and feedback on the manuscript.
R.S. acknowledges support from the Nuclear Physics from Multi-Messenger Mergers (NP3M) Focused Research Hub which is funded by the National Science Foundation under Grant Number 21-16686, and from the Laboratory Directed Research and Development program of Los Alamos National Laboratory under project number 20220541ECR.
C.L.A. was supported by the Laboratory Directed Research and Development program of Los Alamos National Laboratory under project number 20230315ER.
I.T. and S.G. were supported by the U.S. Department of Energy, Office of Science, Office of Nuclear Physics, under contract No.~DE-AC52-06NA25396 and by the U.S. Department of Energy, Office of Science, Office of Advanced Scientific Computing Research, Scientific Discovery through Advanced Computing (SciDAC) NUCLEI program.
I.T. was also supported by the Laboratory Directed Research and Development program of Los Alamos National Laboratory under project numbers 20220541ECR and 20230315ER.
P.G. and K.G. were supported by the National Science Foundation CSSI program under award No.~OAC-2004601 (BAND Collaboration).
Computational resources have been provided  by the Los Alamos National Laboratory Institutional Computing Program, which is supported by the U.S. Department of Energy National Nuclear Security Administration under Contract No.~89233218CNA000001, and by the National Energy Research Scientific Computing Center (NERSC), which is supported by the U.S. Department of Energy, Office of Science, under contract No.~DE-AC02-05CH11231.
\end{sloppypar}

\bibliography{bib}

\begin{thebibliography}{10}
\expandafter\ifx\csname url\endcsname\relax
  \def\url#1{\texttt{#1}}\fi
\expandafter\ifx\csname urlprefix\endcsname\relax\def\urlprefix{URL }\fi
\expandafter\ifx\csname href\endcsname\relax
  \def\href#1#2{#2} \def\path#1{#1}\fi

\bibitem{Koehn:2024set}
H.~Koehn, et~al., {From existing and new nuclear and astrophysical constraints to stringent limits on the equation of state of neutron-rich dense matter}, Phys. Rev. X 15~(2) (2025) 021014.
\newblock \href {http://arxiv.org/abs/2402.04172} {\path{arXiv:2402.04172}}, \href {https://doi.org/10.1103/PhysRevX.15.021014} {\path{doi:10.1103/PhysRevX.15.021014}}.

\bibitem{NANOGrav:2019jur}
H.~T. Cromartie, et~al., {Relativistic Shapiro delay measurements of an extremely massive millisecond pulsar}, Nature Astron. 4~(1) (2019) 72--76.
\newblock \href {http://arxiv.org/abs/1904.06759} {\path{arXiv:1904.06759}}, \href {https://doi.org/10.1038/s41550-019-0880-2} {\path{doi:10.1038/s41550-019-0880-2}}.

\bibitem{LIGOScientific:2017ync}
B.~P. Abbott, et~al., {Multi-messenger Observations of a Binary Neutron Star Merger}, Astrophys. J. Lett. 848~(2) (2017) L12.
\newblock \href {http://arxiv.org/abs/1710.05833} {\path{arXiv:1710.05833}}, \href {https://doi.org/10.3847/2041-8213/aa91c9} {\path{doi:10.3847/2041-8213/aa91c9}}.

\bibitem{LIGOScientific:2017vwq}
B.~P. Abbott, et~al., {GW170817: Observation of Gravitational Waves from a Binary Neutron Star Inspiral}, Phys. Rev. Lett. 119~(16) (2017) 161101.
\newblock \href {http://arxiv.org/abs/1710.05832} {\path{arXiv:1710.05832}}, \href {https://doi.org/10.1103/PhysRevLett.119.161101} {\path{doi:10.1103/PhysRevLett.119.161101}}.

\bibitem{Riley:2019yda}
T.~E. Riley, et~al., {A $NICER$ View of PSR J0030+0451: Millisecond Pulsar Parameter Estimation}, Astrophys. J. Lett. 887~(1) (2019) L21.
\newblock \href {http://arxiv.org/abs/1912.05702} {\path{arXiv:1912.05702}}, \href {https://doi.org/10.3847/2041-8213/ab481c} {\path{doi:10.3847/2041-8213/ab481c}}.

\bibitem{Riley:2021pdl}
T.~E. Riley, et~al., {A NICER View of the Massive Pulsar PSR J0740+6620 Informed by Radio Timing and XMM-Newton Spectroscopy}, Astrophys. J. Lett. 918~(2) (2021) L27.
\newblock \href {http://arxiv.org/abs/2105.06980} {\path{arXiv:2105.06980}}, \href {https://doi.org/10.3847/2041-8213/ac0a81} {\path{doi:10.3847/2041-8213/ac0a81}}.

\bibitem{Miller:2019cac}
M.~C. Miller, et~al., {PSR J0030+0451 Mass and Radius from $NICER$ Data and Implications for the Properties of Neutron Star Matter}, Astrophys. J. Lett. 887~(1) (2019) L24.
\newblock \href {http://arxiv.org/abs/1912.05705} {\path{arXiv:1912.05705}}, \href {https://doi.org/10.3847/2041-8213/ab50c5} {\path{doi:10.3847/2041-8213/ab50c5}}.

\bibitem{Miller:2021qha}
M.~C. Miller, et~al., {The Radius of PSR J0740+6620 from NICER and XMM-Newton Data}, Astrophys. J. Lett. 918~(2) (2021) L28.
\newblock \href {http://arxiv.org/abs/2105.06979} {\path{arXiv:2105.06979}}, \href {https://doi.org/10.3847/2041-8213/ac089b} {\path{doi:10.3847/2041-8213/ac089b}}.

\bibitem{Pang:2022rzc}
P.~T.~H. Pang, et~al., {An updated nuclear-physics and multi-messenger astrophysics framework for binary neutron star mergers}, Nature Commun. 14~(1) (2023) 8352.
\newblock \href {http://arxiv.org/abs/2205.08513} {\path{arXiv:2205.08513}}, \href {https://doi.org/10.1038/s41467-023-43932-6} {\path{doi:10.1038/s41467-023-43932-6}}.

\bibitem{Raaijmakers:2019qny}
G.~Raaijmakers, et~al., {A $NICER$ view of PSR J0030+0451: Implications for the dense matter equation of state}, Astrophys. J. Lett. 887~(1) (2019) L22.
\newblock \href {http://arxiv.org/abs/1912.05703} {\path{arXiv:1912.05703}}, \href {https://doi.org/10.3847/2041-8213/ab451a} {\path{doi:10.3847/2041-8213/ab451a}}.

\bibitem{Farr:2021fyc}
J.~N. Farr, Z.~Meisel, A.~W. Steiner, {Decision Theory for the Mass Measurements at the Facility for Rare Isotope Beams} (11 2021).
\newblock \href {http://arxiv.org/abs/2111.11536} {\path{arXiv:2111.11536}}.

\bibitem{phillips2021get}
D.~R. Phillips, R.~J. Furnstahl, U.~Heinz, T.~Maiti, W.~Nazarewicz, F.~M. Nunes, M.~Plumlee, M.~T. Pratola, S.~Pratt, F.~G. Viens, S.~M. Wild, \href{https://doi.org/10.1088/1361-6471/abf1df}{Get on the {BAND} wagon: a {B}ayesian framework for quantifying model uncertainties in nuclear dynamics}, J. Phys. G Nucl. Part. Phys. 48~(7) (2021) 072001.
\newblock \href {https://doi.org/10.1088/1361-6471/abf1df} {\path{doi:10.1088/1361-6471/abf1df}}.
\newline\urlprefix\url{https://doi.org/10.1088/1361-6471/abf1df}

\bibitem{Carlson:2014vla}
J.~Carlson, S.~Gandolfi, F.~Pederiva, S.~C. Pieper, R.~Schiavilla, K.~E. Schmidt, R.~B. Wiringa, {Quantum Monte Carlo methods for nuclear physics}, Rev. Mod. Phys. 87 (2015) 1067.
\newblock \href {http://arxiv.org/abs/1412.3081} {\path{arXiv:1412.3081}}, \href {https://doi.org/10.1103/RevModPhys.87.1067} {\path{doi:10.1103/RevModPhys.87.1067}}.

\bibitem{Schmidt:1999lik}
K.~E. Schmidt, S.~Fantoni, {A quantum Monte Carlo method for nucleon systems}, Phys. Lett. B 446 (1999) 99--103.
\newblock \href {https://doi.org/10.1016/S0370-2693(98)01522-6} {\path{doi:10.1016/S0370-2693(98)01522-6}}.

\bibitem{Gezerlis:2013ipa}
A.~Gezerlis, I.~Tews, E.~Epelbaum, S.~Gandolfi, K.~Hebeler, A.~Nogga, A.~Schwenk, {Quantum Monte Carlo Calculations with Chiral Effective Field Theory Interactions}, Phys. Rev. Lett. 111~(3) (2013) 032501.
\newblock \href {http://arxiv.org/abs/1303.6243} {\path{arXiv:1303.6243}}, \href {https://doi.org/10.1103/PhysRevLett.111.032501} {\path{doi:10.1103/PhysRevLett.111.032501}}.

\bibitem{Gezerlis:2014zia}
A.~Gezerlis, I.~Tews, E.~Epelbaum, M.~Freunek, S.~Gandolfi, K.~Hebeler, A.~Nogga, A.~Schwenk, {Local chiral effective field theory interactions and quantum Monte Carlo applications}, Phys. Rev. C 90~(5) (2014) 054323.
\newblock \href {http://arxiv.org/abs/1406.0454} {\path{arXiv:1406.0454}}, \href {https://doi.org/10.1103/PhysRevC.90.054323} {\path{doi:10.1103/PhysRevC.90.054323}}.

\bibitem{Tews:2015ufa}
I.~Tews, S.~Gandolfi, A.~Gezerlis, A.~Schwenk, {Quantum Monte Carlo calculations of neutron matter with chiral three-body forces}, Phys. Rev. C 93~(2) (2016) 024305.
\newblock \href {http://arxiv.org/abs/1507.05561} {\path{arXiv:1507.05561}}, \href {https://doi.org/10.1103/PhysRevC.93.024305} {\path{doi:10.1103/PhysRevC.93.024305}}.

\bibitem{Lynn:2015jua}
J.~E. Lynn, I.~Tews, J.~Carlson, S.~Gandolfi, A.~Gezerlis, K.~E. Schmidt, A.~Schwenk, {Chiral Three-Nucleon Interactions in Light Nuclei, Neutron-$\alpha$ Scattering, and Neutron Matter}, Phys. Rev. Lett. 116~(6) (2016) 062501.
\newblock \href {http://arxiv.org/abs/1509.03470} {\path{arXiv:1509.03470}}, \href {https://doi.org/10.1103/PhysRevLett.116.062501} {\path{doi:10.1103/PhysRevLett.116.062501}}.

\bibitem{Somasundaram:2023sup}
R.~Somasundaram, J.~E. Lynn, L.~Huth, A.~Schwenk, I.~Tews, {Maximally local two-nucleon interactions at N3LO in \ensuremath{\Delta}-less chiral effective field theory}, Phys. Rev. C 109~(3) (2024) 034005.
\newblock \href {http://arxiv.org/abs/2306.13579} {\path{arXiv:2306.13579}}, \href {https://doi.org/10.1103/PhysRevC.109.034005} {\path{doi:10.1103/PhysRevC.109.034005}}.

\bibitem{Lonardoni:2017hgs}
D.~Lonardoni, J.~Carlson, S.~Gandolfi, J.~E. Lynn, K.~E. Schmidt, A.~Schwenk, X.~Wang, {Properties of nuclei up to $A=16$ using local chiral interactions}, Phys. Rev. Lett. 120~(12) (2018) 122502.
\newblock \href {http://arxiv.org/abs/1709.09143} {\path{arXiv:1709.09143}}, \href {https://doi.org/10.1103/PhysRevLett.120.122502} {\path{doi:10.1103/PhysRevLett.120.122502}}.

\bibitem{Bonilla:2022rph}
E.~Bonilla, P.~Giuliani, K.~Godbey, D.~Lee, {Training and projecting: A reduced basis method emulator for many-body physics}, Phys. Rev. C 106~(5) (2022) 054322.
\newblock \href {http://arxiv.org/abs/2203.05284} {\path{arXiv:2203.05284}}, \href {https://doi.org/10.1103/PhysRevC.106.054322} {\path{doi:10.1103/PhysRevC.106.054322}}.

\bibitem{Melendez:2022kid}
J.~A. Melendez, C.~Drischler, R.~J. Furnstahl, A.~J. Garcia, X.~Zhang, {Model reduction methods for nuclear emulators}, J. Phys. G 49~(10) (2022) 102001.
\newblock \href {http://arxiv.org/abs/2203.05528} {\path{arXiv:2203.05528}}, \href {https://doi.org/10.1088/1361-6471/ac83dd} {\path{doi:10.1088/1361-6471/ac83dd}}.

\bibitem{Duguet:2023wuh}
T.~Duguet, A.~Ekstr\"om, R.~J. Furnstahl, S.~K\"onig, D.~Lee, {Colloquium: Eigenvector continuation and projection-based emulators}, Rev. Mod. Phys. 96~(3) (2024) 031002.
\newblock \href {http://arxiv.org/abs/2310.19419} {\path{arXiv:2310.19419}}, \href {https://doi.org/10.1103/RevModPhys.96.031002} {\path{doi:10.1103/RevModPhys.96.031002}}.

\bibitem{Frame:2017fah}
D.~Frame, R.~He, I.~Ipsen, D.~Lee, D.~Lee, E.~Rrapaj, {Eigenvector continuation with subspace learning}, Phys. Rev. Lett. 121~(3) (2018) 032501.
\newblock \href {http://arxiv.org/abs/1711.07090} {\path{arXiv:1711.07090}}, \href {https://doi.org/10.1103/PhysRevLett.121.032501} {\path{doi:10.1103/PhysRevLett.121.032501}}.

\bibitem{Konig:2019adq}
S.~K\"onig, A.~Ekstr\"om, K.~Hebeler, D.~Lee, A.~Schwenk, {Eigenvector Continuation as an Efficient and Accurate Emulator for Uncertainty Quantification}, Phys. Lett. B 810 (2020) 135814.
\newblock \href {http://arxiv.org/abs/1909.08446} {\path{arXiv:1909.08446}}, \href {https://doi.org/10.1016/j.physletb.2020.135814} {\path{doi:10.1016/j.physletb.2020.135814}}.

\bibitem{Lay:2023boz}
D.~Lay, E.~Flynn, S.~A. Giuliani, W.~Nazarewicz, L.~Neufcourt, {Neural network emulation of spontaneous fission}, Phys. Rev. C 109~(4) (2024) 044305.
\newblock \href {http://arxiv.org/abs/2310.01608} {\path{arXiv:2310.01608}}, \href {https://doi.org/10.1103/PhysRevC.109.044305} {\path{doi:10.1103/PhysRevC.109.044305}}.

\bibitem{surer2022uncertainty}
{\"O}.~S{\"u}rer, F.~M. Nunes, M.~Plumlee, S.~M. Wild, Uncertainty quantification in breakup reactions, Physical Review C 106~(2) (2022) 024607.

\bibitem{rose2024}
D.~Odell, P.~Giuliani, K.~Beyer, M.~Catacora-Rios, M.~Y.-H. Chan, E.~Bonilla, R.~J. Furnstahl, K.~Godbey, F.~M. Nunes, \href{https://link.aps.org/doi/10.1103/PhysRevC.109.044612}{Rose: A reduced-order scattering emulator for optical models}, Phys. Rev. C 109 (2024) 044612.
\newblock \href {https://doi.org/10.1103/PhysRevC.109.044612} {\path{doi:10.1103/PhysRevC.109.044612}}.
\newline\urlprefix\url{https://link.aps.org/doi/10.1103/PhysRevC.109.044612}

\bibitem{drischler2023buqeye}
C.~Drischler, J.~Melendez, R.~Furnstahl, A.~Garcia, X.~Zhang, Buqeye guide to projection-based emulators in nuclear physics, Frontiers in Physics 10 (2023) 1092931.

\bibitem{GP_book}
C.~E. Rasmussen, C.~K.~I. Williams, \href{https://doi.org/10.7551/mitpress/3206.001.0001}{{Gaussian Processes for Machine Learning}}, The MIT Press, 2005.
\newblock \href {https://doi.org/10.7551/mitpress/3206.001.0001} {\path{doi:10.7551/mitpress/3206.001.0001}}.
\newline\urlprefix\url{https://doi.org/10.7551/mitpress/3206.001.0001}

\bibitem{hesthaven2016certified}
J.~S. Hesthaven, G.~Rozza, B.~Stamm, et~al., {Certified Reduced Basis Methods for Parametrized Partial Differential Equations}, Vol. 590, Springer, 2016.
\newblock \href {https://doi.org/https://doi.org/10.1007/978-3-319-22470-1} {\path{doi:https://doi.org/10.1007/978-3-319-22470-1}}.

\bibitem{PG_book}
J.~N. Reddy, An introduction to the finite element method, 3rd Edition, McGraw-Hill Higher Education, New York, NY, 2006.

\bibitem{J_1984}
C.~A. Fletcher, Computational galerkin methods, Springer, 1984.
\newblock \href {https://doi.org/https://doi.org/10.1007/978-3-642-85949-6} {\path{doi:https://doi.org/10.1007/978-3-642-85949-6}}.

\bibitem{brunton2019data}
S.~L. Brunton, J.~N. Kutz, Data-driven science and engineering: Machine learning, dynamical systems, and control, Cambridge University Press, 2019.
\newblock \href {https://doi.org/https://doi.org/10.1017/9781108380690} {\path{doi:https://doi.org/10.1017/9781108380690}}.

\bibitem{Ekstrom:2019lss}
A.~Ekstr\"om, G.~Hagen, {Global sensitivity analysis of bulk properties of an atomic nucleus}, Phys. Rev. Lett. 123~(25) (2019) 252501.
\newblock \href {http://arxiv.org/abs/1910.02922} {\path{arXiv:1910.02922}}, \href {https://doi.org/10.1103/PhysRevLett.123.252501} {\path{doi:10.1103/PhysRevLett.123.252501}}.

\bibitem{Jiang:2022tzf}
W.~G. Jiang, C.~Forss\'en, T.~Dj\"arv, G.~Hagen, {Nuclear-matter saturation and symmetry energy within \ensuremath{\Delta}-full chiral effective field theory}, Phys. Rev. C 109~(6) (2024) L061302.
\newblock \href {http://arxiv.org/abs/2212.13203} {\path{arXiv:2212.13203}}, \href {https://doi.org/10.1103/PhysRevC.109.L061302} {\path{doi:10.1103/PhysRevC.109.L061302}}.

\bibitem{Jiang:2022oba}
W.~G. Jiang, C.~Forss\'en, T.~Dj\"arv, G.~Hagen, {Emulating ab initio computations of infinite nucleonic matter}, Phys. Rev. C 109~(6) (2024) 064314.
\newblock \href {http://arxiv.org/abs/2212.13216} {\path{arXiv:2212.13216}}, \href {https://doi.org/10.1103/PhysRevC.109.064314} {\path{doi:10.1103/PhysRevC.109.064314}}.

\bibitem{Frame:2019jsw}
D.~K. Frame, {Ab initio simulations of light nuclear systems using eigenvector continuation and auxiliary field Monte Carlo}, Ph.D. thesis, Michigan State U., Michigan State U. (2019).
\newblock \href {http://arxiv.org/abs/1905.02782} {\path{arXiv:1905.02782}}, \href {https://doi.org/10.25335/rjj4-8347} {\path{doi:10.25335/rjj4-8347}}.

\bibitem{Sarkar:2023qjn}
A.~Sarkar, D.~Lee, U.-G. Mei\ss{}ner, {Floating Block Method for Quantum Monte~Carlo Simulations}, Phys. Rev. Lett. 131~(24) (2023) 242503.
\newblock \href {http://arxiv.org/abs/2306.11439} {\path{arXiv:2306.11439}}, \href {https://doi.org/10.1103/PhysRevLett.131.242503} {\path{doi:10.1103/PhysRevLett.131.242503}}.

\bibitem{Cook:2024toj}
P.~Cook, D.~Jammooa, M.~Hjorth-Jensen, D.~D. Lee, D.~Lee, {Parametric Matrix Models} (1 2024).
\newblock \href {http://arxiv.org/abs/2401.11694} {\path{arXiv:2401.11694}}.

\bibitem{Lynn:2019rdt}
J.~E. Lynn, I.~Tews, S.~Gandolfi, A.~Lovato, {Quantum Monte Carlo Methods in Nuclear Physics: Recent Advances}, Ann. Rev. Nucl. Part. Sci. 69 (2019) 279--305.
\newblock \href {http://arxiv.org/abs/1901.04868} {\path{arXiv:1901.04868}}, \href {https://doi.org/10.1146/annurev-nucl-101918-023600} {\path{doi:10.1146/annurev-nucl-101918-023600}}.

\bibitem{Epelbaum:2004fk}
E.~Epelbaum, W.~Glockle, U.-G. Meissner, {The Two-nucleon system at next-to-next-to-next-to-leading order}, Nucl. Phys. A 747 (2005) 362--424.
\newblock \href {http://arxiv.org/abs/nucl-th/0405048} {\path{arXiv:nucl-th/0405048}}, \href {https://doi.org/10.1016/j.nuclphysa.2004.09.107} {\path{doi:10.1016/j.nuclphysa.2004.09.107}}.

\bibitem{quarteroni2011certified}
A.~Quarteroni, G.~Rozza, A.~Manzoni, Certified reduced basis approximation for parametrized partial differential equations and applications, J. Math. Ind. 1~(1) (2011) 1--49.
\newblock \href {https://doi.org/https://doi.org/10.1186/2190-5983-1-3} {\path{doi:https://doi.org/10.1186/2190-5983-1-3}}.

\bibitem{drnuclear}
K.~Godbey, P.~Giuliani, E.~Bonilla, E.~Flynn, D.~Odell, K.~Beyer, D.~Lay, D.~Figueroa, R.~Garg, M.~Campbell, Dimensionality reduction in nuclear physics, \url{https://dr.ascsn.net/blackbox_galerkin/Introduction.html}.

\bibitem{wales1997global}
D.~J. Wales, J.~P. Doye, Global optimization by basin-hopping and the lowest energy structures of lennard-jones clusters containing up to 110 atoms, The Journal of Physical Chemistry A 101~(28) (1997) 5111--5116.

\bibitem{kullback1951information}
S.~Kullback, R.~A. Leibler, On information and sufficiency, The annals of mathematical statistics 22~(1) (1951) 79--86.

\bibitem{Armstrong:2025tza}
C.~L. Armstrong, P.~Giuliani, K.~Godbey, R.~Somasundaram, I.~Tews, {Emulators for scarce and noisy data II: Application to auxiliary-field diffusion Monte Carlo for neutron matter} (2 2025).
\newblock \href {http://arxiv.org/abs/2502.03680} {\path{arXiv:2502.03680}}.

\end{thebibliography}

\end{document}